\documentclass[10pt,twocolumn,twoside,conference]{IEEEtran}
\IEEEoverridecommandlockouts                              
\usepackage{graphicx} 
\usepackage{subcaption}
\usepackage{dblfloatfix}
\usepackage{amsmath,amsthm,amssymb,amsfonts} 
\usepackage{cite}
\usepackage{verbatim}
\usepackage{algorithm}
\usepackage{subfiles}
\usepackage{setspace}
\usepackage{float}
\usepackage{wrapfig}
\usepackage{color}

\allowdisplaybreaks

\providecommand{\half}{\frac{1}{2}}

\providecommand{\vsn}[1][n]{\mathcal{S}_{#1}}
\providecommand{\vc}{\mathcal{C}}
\providecommand{\ct}{\Gamma}

\providecommand{\rhovect}{\boldsymbol{\rho}}

\providecommand{\van}[1][n]{\boldsymbol{\alpha}^{(#1)}}
\providecommand{\vbn}[1][n]{\boldsymbol{\beta}^{(#1)}}
\providecommand{\vrn}[1][n]{\mathbf{r}^{(#1)}}

\providecommand{\vY}{\mathbf{Y}}
\providecommand{\vy}{\mathbf{y}}
\providecommand{\vYn}[1][n]{\vY^{(#1)}}
\providecommand{\vyn}[1][n]{\vy^{(#1)}}
\providecommand{\vYnk}[1][n]{ \vY^{(#1)}_{-k} }

\providecommand{\vzn}[1][n]{\boldsymbol{0}^{(#1)}}
\providecommand{\vin}[1][n]{\boldsymbol{1}^{(#1)}}

\providecommand{\cta}[1][{\pi^{\ast}}]{\ct_0^{#1}}
\providecommand{\ctb}[1][{\pi^{\ast}}]{\ct_1^{#1}}

\providecommand{\ctbk}[2][\pi^{\ast}]{\ct^{#1}_{1,Y_k = #2}}

\providecommand{\proda}{\tilde{P}_{\boldsymbol{\alpha}}}
\providecommand{\prodb}{\tilde{P}_{\boldsymbol{\beta}}}
\providecommand{\prodab}{P_{\boldsymbol{\alpha-\beta}}}

\providecommand{\proddiv}{P_{\boldsymbol{\alpha/\beta}}}

\providecommand{\half}{\frac{1}{2}}
\providecommand{\smallratio}{\frac{\rho_1}{\rho_0}}

\providecommand{\indicator}{\mathbf{1}}

\providecommand{\ct}{\mathcal{T}}

\providecommand{\cch}{\mathcal{C}}

\providecommand{\vsn}[1][n]{\mathbf{\cs}_{#1}}
\providecommand{\vCn}[1][n]{\mathbf{\cch}^{#1}}

\providecommand{\nat}{\mathbb{N}}

\providecommand{\vYn}[1][n]{\mathbf{Y}^{\left(#1\right)}}
\providecommand{\vyn}[1][n]{\mathbf{y}^{\left(#1\right)}}
\providecommand{\vy}{\mathbf{y}}

\providecommand{\van}[1][n]{\boldsymbol{\alpha}^{\left(#1\right)}}
\providecommand{\vbn}[1][n]{\boldsymbol{\beta}^{\left(#1\right)}}
\providecommand{\vrn}[1][n]{\mathbf{r}\small(\rho\small)^{(#1)}}

\providecommand{\vin}[1][n]{\mathbf{e}^{(#1)}}
\providecommand{\vzn}[1][n]{\mathbf{0}^{(#1)}}

\providecommand{\rhovect}{\boldsymbol{\rho}}

\providecommand{\proda}[1][n]{\widetilde{\mathbb{P}}_{\alpha} \left( \vyn[#1] \right)}
\providecommand{\prodb}[1][n]{\widetilde{\mathbb{P}}_{\beta} \left( \vyn[#1] \right)}

\providecommand{\modfloor}[1]{\left \lfloor #1 \right \rfloor}
\providecommand{\modceil}[1]{\left \lceil #1 \right \rceil}

\newtheorem{mythm}{Theorem}
\newtheorem{mycorr}{Corollary}
\newtheorem{myclaim}{Claim}

\newtheorem{mydef}{Definition}

\DeclareGraphicsExtensions{.pdf,.png,.jpg,.eps}

\def\BibTeX{{\rm B\kern-.05em{\sc i\kern-.025em b}\kern-.08em
    T\kern-.1667em\lower.7ex\hbox{E}\kern-.125emX}}

\begin{document}
 
\begin{titlepage}
$\copyright$ 2020 IEEE. Personal use of this material is permitted. Permission from IEEE must be
obtained for all other uses, in any current or future media, including
reprinting/republishing this material for advertising or promotional purposes, creating new
collective works, for resale or redistribution to servers or lists, or reuse of any copyrighted
component of this work in other works.
\end{titlepage}

\title{\LARGE \bf Optimal Decisions of a Rational Agent \linebreak in the Presence of Biased Information Providers}



\author{\IEEEauthorblockN{Himaja Kesavareddigari}
\IEEEauthorblockA{\textit{ECE Dept., The Ohio State University} \\
kesavareddigari.1@osu.edu}
\and
\IEEEauthorblockN{Atilla Eryilmaz}
\IEEEauthorblockA{\textit{ECE Dept., The Ohio State University} \\
eryilmaz.2@osu.edu}
\thanks{This work was supported primarily by the ONR Grant N00014-19-1-2621, and in part by the NSF grants: CNS-NeTS-1514260, CNS-NeTS-1717045, CMMISMOR-1562065, CNS-ICN-WEN-1719371, and CNS-SpecEES-18243; and the DTRA grant HDTRA1-18-1-0050.
}
}




\maketitle

\begin{abstract}

We consider information networks whereby multiple biased-information-providers (BIPs), e.g., media outlets/social network users/sensors, share reports of events with rational-information-consumers (RICs). Making the reasonable abstraction that an event can be reported as an answer to a logical statement, we model the input-output behavior of each BIP as a binary channel. For various reasons, some BIPs might share incorrect reports of the event. Moreover, each BIP is: `biased' if it favors one of the two outcomes while reporting, or `unbiased' if it favors neither outcome. Such biases occur in information/social networks due to differences in users' characteristics/worldviews.

We study the impact of the BIPs' biases on an RIC's choices while deducing the true information. Our work reveals that a ``graph-blind" RIC looking for $n$ BIPs among its neighbors, \emph{acts peculiarly in order to minimize its probability of making an error while deducing the true information}. First, we establish the counter-intuitive fact that the RIC's expected error is minimized by \emph{choosing BIPs that are fully-biased against the a-priori likely event}. Then, we study the gains that fully-biased BIPs provide over unbiased BIPs when the error rates of their binary channels are equalized, for fair comparison, at some $r>0$. Specifically, the unbiased-to-fully-biased ratio of the RIC's expected error probabilities grows exponentially with the exponent $\frac{n}{2}\ln\left(4\rho_0^2\left(\frac{1}{r}-1\right)\right)$, where $\rho_0$ is the event's prior probability of being $0$. This shows not only that fully-biased BIPs are preferable to unbiased or heterogeneously-biased BIPs, but also that the gains can be substantial for small $r$.

\end{abstract}



\section{Introduction}
\label{sec:intro}
%
%
We consider an information network connecting users that may be devices with limited resources or human, and are biased-information-providers (BIPs) and/or rational-information-consumers (RICs). We assume that reports of a single event that occurred outside this network are propagated as a true(1) or false(0) answer to a logical statement. While our network model and analysis are applicable to any information network, we were primarily motivated by the unique qualities and constraints presented by modern-day social networks.


That is, a single bit of $0/1$ information from a source is being transferred via BIPs to the RICs. However, the BIPs may report incorrectly for a variety of reasons; we model a BIP's reporting behavior as a binary channel to depict the errors in its reporting of the input bit. In particular, each BIP can possess a \emph{bias} favoring either the $0$ bit or the $1$ bit in its reporting. Since RICs (e.g. low-memory devices, humans) might not know the network graph except for a list of their own neighbors, we can reasonably assume that they are \emph{graph-blind}. Therefore, when trying to minimize its consumption of false information, \textit{a ``graph-blind" RIC will have to assume that all of its BIPs are acting independently}. This assumption of independence is a reasonable approximation of the typical behavior of RICs that are either human or low-memory devices. 

The goal of this work is to perform a careful study of the impact of the information providers' biases on the choices of an RIC that is attempting to accurately detect the original information. Given that BIPs are inevitably unreliable \textit{and} biased, we are especially interested in unearthing ``how" the BIPs' biases impact the RIC, and by ``how much". 
In online social networks (OSNs), $0/1$ might represent contradicting depictions/viewpoints of a current event. For a particular social network user, its trusted media outlets and friends are the BIPs.
A BIP's favorable opinion of arguments/evidence supporting one viewpoint as opposed to the other, might color its reports of the original/source information.

Even though the design is inspired and constrained by the particular nature of interactions on OSNs, this setting finds application in other information networks. For example, sensor networks where the devices have unequal false alarm and misdetection probabilities. Herein, the $0/1$ information at the source can represent the absence/occurrence of a sensor-triggering event. Akin to the OSN users described earlier, the sensors might be graph-blind, unreliable, and might also have asymmetric sensitivities to the $0/1$ triggering event (i.e., causing more false alarms than misdetections and vice versa).


Our problem statement bears comparison to these topics in literature: Containment of Misinformation, Information Theory, Social Sensing, and Information Fusion.

Influence maximization and containment of misinformation on social networks, by placing influential and protector nodes strategically, is studied in literature 
(e.g.,~\cite{bharathi2007competitive,nguyen2013analysis}). Our approach differs from these works in that we are not interested in the NP-hard problem of finding the most influential nodes to target, in a network. Instead, we are interested in the decisions of any rational-information-consumer choosing from a (possibly, large) selection of biased-information-providers.


Reliable transfer of information over unreliable binary channels is widely studied in information theory (e.g.~\cite{cover2012elements,shannon1967lower,silverman1955binary,li2016survey}). Wherein, unlike the problem statement of this paper, the BIPs might choose to encode the information or transmit information about a block of $n_e$ events at the same time, to increase the rate of reliable information transfer (a.k.a. channel capacity). However, in OSNs and some sensor networks, the BIPs lack accountability and it is more practical to transmit information about each event separately and without delay.
Here, BIPs are error-prone, biased, \emph{abundantly available and act promptly}. This requires methods for reliable information transfer when \emph{only one event} is processed at a time.


Social sensing literature studied the problem of the true value of a solitary binary quantity based on data arriving from multiple data sources of unknown credibility 
(e.g.,\cite{wang2013credibility,dong2009integrating,yin2008truth,wang2013recursive,liu2015social}). These works, however, are focused on identifying duplicates and dependencies in incoming datasets and  estimating the credibility of sources to maximize the probability of discovering the ground truth. As such, they do not consider the impact of information-providers' biases.


Information fusion studies the detection of ground truth in information (mostly, sensor) networks 
(e.g.~\cite{chair1986optimal,samarasooriya1996sequential,varshney1996distributed,brooks2003distributed,niu2004decision,nakamura2007information,abu2018localised,sriranga2018energy,vempaty2018experiments}). These algorithms assume channel and noise characteristics, and may process signals sequentially, in the order of a sensor's reliability, accounting for channel fading, sensor's distance to its measuring target, etc. 
Our investigation contributes to these efforts, with a unique perspective on how, from abundantly many BIPs, \emph{a graph-blind RIC will choose BIPs that are optimally-biased, given their individual error rate.}

While intuition might dictate that an unbiased BIP is a better choice than a biased one, we find that a \emph{rational}-information-consumer will choose the opposite.
%
Our main contributions can be listed as follows:
\begin{list}{$\bullet$}{\leftmargin=1em \itemindent=0em \parskip=-1em \parsep=-1pt}
    \item When choosing independent BIPs to report a $0/1$ event: we find that an RIC is best served by the BIPs that are fully biased against the a-priori likely event. Using this optimal choice of BIPs, the RIC will deduce that the a-priori likely event is true unless all BIPs report that the a-priori unlikely event is true (cf. Corollary~\ref{thm:extremebias}). Further, if it is not possible to obtain BIPs that are fully-biased against the a-priori likely event, then the set of $n$ BIPs that will best serve the RIC are still guaranteed to be maximally-biased (cf. Theorem~\ref{thm:concave}).
    \item Our analysis shows that when a system of unbiased BIPs (that do not favor either outcome) is replaced by a system of fully-biased BIPs (favoring the same outcome), the gain (in terms of probability of error) rises as an exponential function of a positive exponent (cf. Theorem~\ref{thm:gains}) that decreases with the BIPs' error rate and increases linearly with $n$.
\end{list}
We establish that an RIC acts according to a counter-intuitive, but very tractable rule. As such, we consider RICs with limited knowledge of the network, model their surprising behaviors and their perceived benefits of fully-biased information-providers. In future work, we aim to model RICs that exhibit more network awareness and, thus, enable novel RIC-centric frameworks with mechanisms to reduce misinformation.

The rest of the paper is organized as follows. In Section~\ref{sec:model}, we introduce the model of the multi-BIP system of interest. In Section~\ref{sec:main}, we discuss the characteristics of the optimal decision rule and state the main results from our analysis of biases and their impact on a graph-blind RIC's choices. In Section~\ref{sec:lowerbound}, we provide the proofs of the results stated in Section~\ref{sec:main}. Finally, in Section~\ref{sec:exp}, we numerically support our findings on the gains that an RIC expects to obtain by choosing BIPs that are fully-biased against the a-priori likely outcome.

\section{System Model} \label{sec:model}
\begin{figure}[!t]
    \centering
    \begin{subfigure}[b]{0.18\textwidth}
        \includegraphics[width=\textwidth,keepaspectratio]{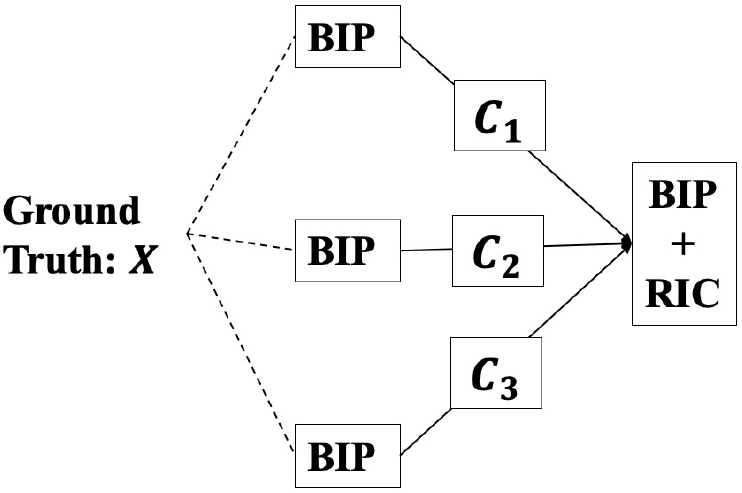}
        \caption{A graph-blind RIC's perspective of the network.}
        \label{fig:indp_channels}
    \end{subfigure}
    \begin{subfigure}[b]{0.3\textwidth}
        \includegraphics[width=\textwidth,keepaspectratio]{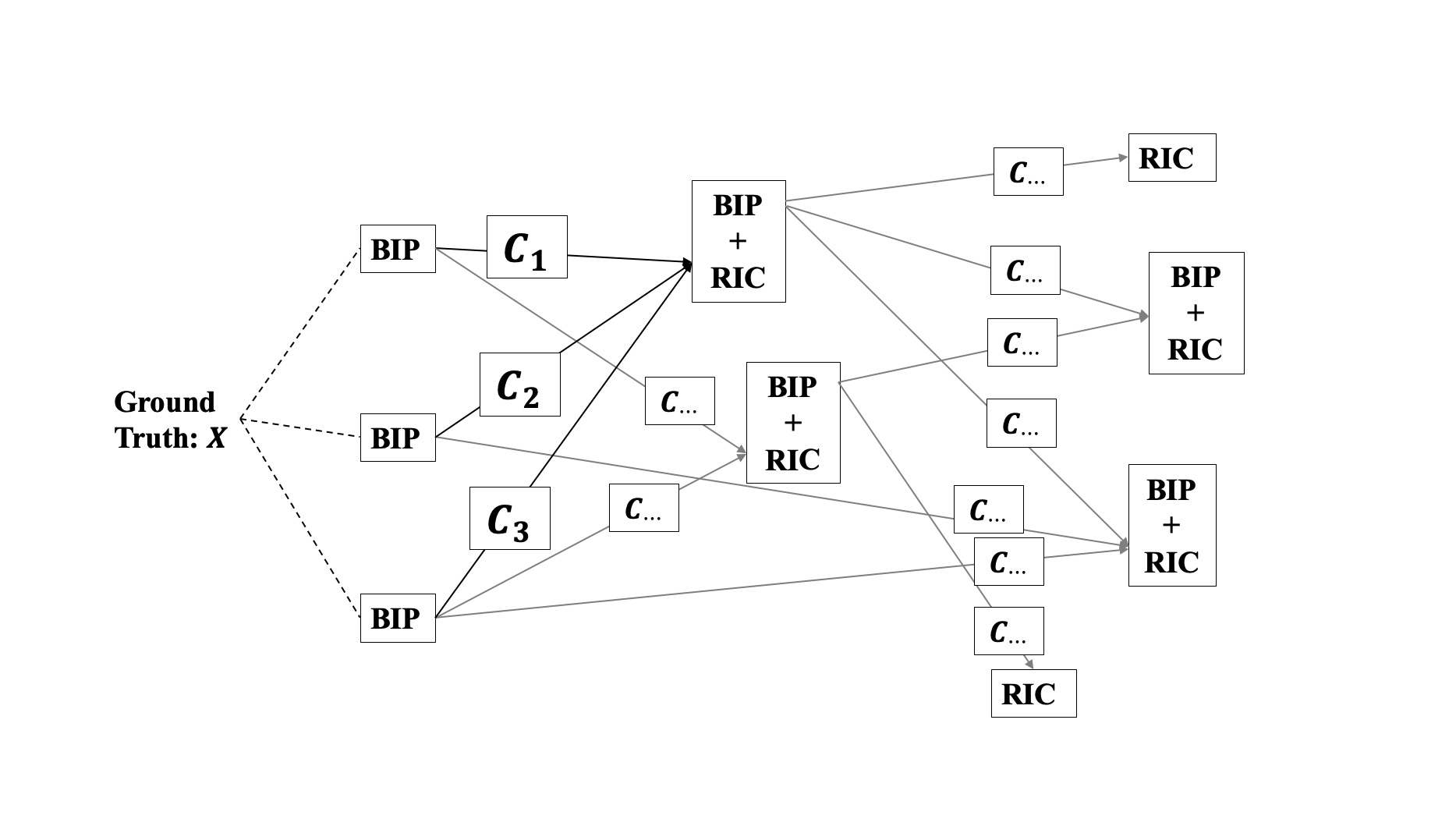}
        \caption{Network of BIPs and RICs.}
        \label{fig:social_network}
    \end{subfigure}
    \caption{Information Transfer in a Network.}
    \label{fig:info_osn}
\end{figure}

\noindent \textbf{Network of Information Providers and Consumers:} In this work, we are interested in understanding how the notion of bias (Defn~\ref{def:bias}) affects the behaviors of a ``graph-blind" RIC whose perspective of the graph is limited (cf. Fig~\ref{fig:indp_channels}). Usually, information networks are more sophisticated, and connect users that are either RICs or BIPs or both (cf. Fig~\ref{fig:social_network}).


It is natural to assume that an RIC would have limited knowledge of the global network and, therefore, will \emph{perceive} its neighboring BIPs as \emph{independent agents who are directly accessing the ground truth}. 
We also reason that an RIC can easily quantify the biases of its information-providers (by using the information-provider's history of reporting, for example). \textit{Our goal is to identify an RIC's perceived-best strategy for choosing $n$ BIPs from a potentially large set of BIPs, assuming the RIC is ``graph-blind"~Fig.~\ref{fig:indp_channels}.}
%

\begin{figure}[!t]
    \centering
    \begin{subfigure}[b]{0.3\textwidth}
        \includegraphics[width=\textwidth,keepaspectratio]{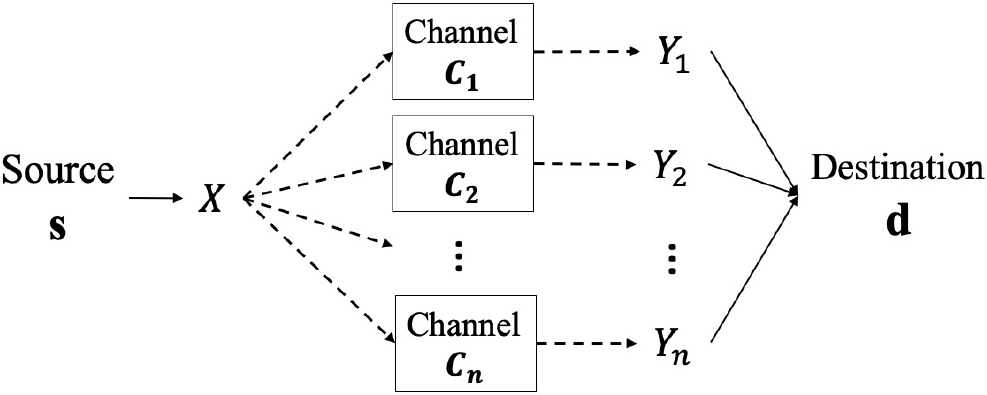}
        \caption{System of $n$ independent BIPs modeled as $n$ parallel channels connecting a source $\mathbf{s}$ and destination $\mathbf{d}$.}
        \label{fig:parallel_channels}
    \end{subfigure}
    \begin{subfigure}[b]{0.18\textwidth}
        \includegraphics[width=\textwidth,keepaspectratio]{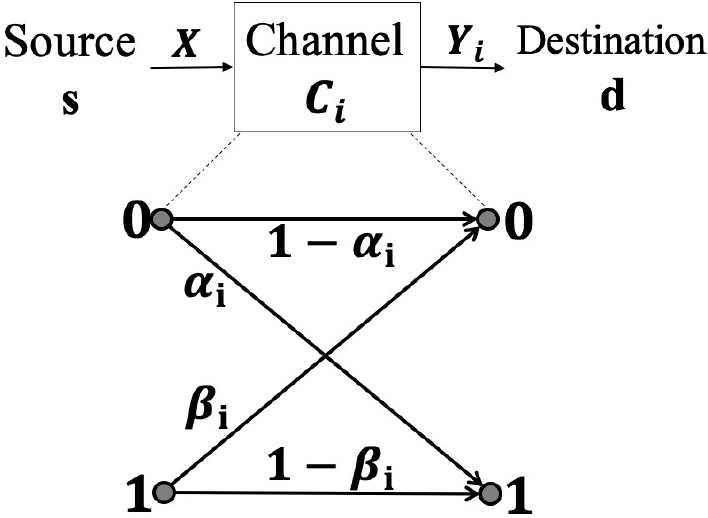}
        \caption{Input-output behavior of the $i$-th BIP modeled by channel $C_i$.}
        \label{fig:channel}
    \end{subfigure}
    \caption{Information Transfer through $n$ independent BIPs.}
    \label{fig:info_channel}
\end{figure}

\noindent \textbf{Multi-Channel Communication Model:} The system in Fig~(\ref{fig:indp_channels}) can be modeled as shown in Fig~(\ref{fig:parallel_channels}), where the information from a source $\mathbf{s}$ arrives at an RIC at destination $\mathbf{d}$ through $n$ BIPs, whose reporting behavior can be modeled as \emph{$n$ parallel, independent binary channels}. In the rest of the paper, we will use the terms `BIP' and `channel" interchangeably.

%
%
Let $\{X(t)\}_t$ be an information stream, where $X(t) \in \{0,1\}$ is a binary random variable representing the information available at the source $\mathbf{s}$ at time $t$. We assume that $X(t)$ is independent of ${X(t'), \forall t' \neq t}$ and can be examined individually. 
We will fix the time instant $t$ and fix $X := X(t)$.

We assume that the prior probability distribution of $X$ is given by $\rhovect = (\rho_0, \rho_1)$ such that $\rho_0 = P(X = 0)$ and $\rho_1 = P(X = 1)$. Without loss of generality, we assume $\rho_0 \geq \rho_1$.

Given $X$, we denote the information received by an RIC through its BIPs' channels, using the binary random vector $\vYn = \{Y_i\}_{i = 1}^n$. The random variable $Y_i$ corresponds to the channel $C_i$, $\forall i \in \{1, \cdots, n\}$.
We represent the behavior of the BIP represented by the binary channel $C_i$ in Fig.~\ref{fig:channel}, where $\alpha_i = P(Y_i = 1 | X = 0) \in [0,1]$ and $\beta_i = P(Y_i = 0 | X = 1) \in [0,1]$.
%
%
%
%
Let the properties of the channel $C_i$ be given by ${\cch_i := (\alpha_i, \beta_i)}$ (cf. Fig.~\ref{fig:channel}). We denote the parameters of the $n$ independent channels and their collective system as ${\vCn := \{n,\van,\vbn\}}$ and ${\vsn := \{n, \rho_0, \van, \vbn\}}$, respectively. Here, ${\van = \{\alpha_i\}_{i = 1}^n}$ and ${\vbn = \{\beta_i\}_{i = 1}^n}$.

\noindent \textbf{Decision Policy:} We are interested in the value of $X \in \{0,1\}$ that is more likely to generate $\vyn = \left\{ y_i \right\}_{i=1}^n \in \{0,1\}^n$, as a realization of the random vector $\vYn$, when applied to $n$ independent BIPs.
In other words, we are interested in a decision policy $\pi$: $\{0,1\}^n \mapsto \{0,1\}$ that achieves the smallest probability of error on $\vsn$. We denote the expected probability of error of decision policy $\pi$ in the system $\vsn$ by $P_e^{\pi}(\vsn)$ and define it as follows.
\begin{align}
P_e^{\pi}(\vsn) 
&= \underset{x \in \{0,1\}}{\sum} \rho_x P \left\{ \pi \left( \vYn \right) \neq X | X = x \right\}. \label{eq:errprob}
\end{align}

Let $\Pi^{\ast} \left(\vsn\right)$ be the set of error-optimal decision policies in a system with parameters $\vsn$. Therefore, for an error-optimal decision policy $\pi^{\ast} \in \Pi^{\ast} \left(\vsn\right),$ 
\begin{align}
& \pi^{\ast} \left( \vyn \right) = \underset{x \in \left\{0,1\right\}}{\arg \max} \; P \left\{X = x; \vYn = \vyn \middle| \vsn\right\}, \notag\\
& P_e^{\pi^{\ast}} \left( \vsn \right) = \underset{\vyn}{\sum} \underset{x \in \left\{0,1\right\}}{\min} \rho_x P \left\{\vYn = \vyn \middle| X = x\right\}.\label{eq:minerrprob}
\end{align}
%
%
\noindent \textbf{Channel Bias:} The parameters $(\alpha_i,\beta_i)$ capture the \textit{bias} of the BIP modeled by $C_i$ (see Fig.~\ref{fig:channel}). So, $\alpha_i > \beta_i$ implies that $C_i$ changes a $0$ input to a $1$ output \emph{more readily} than it changes a $1$ input to a $0$ output. And if $\alpha_i < \beta_i$, the opposite holds true.
We are interested in the effect of the biases $(\van, \vbn)$ on the least probability of error that the RIC hopes to achieve while assuming its $n$ BIPs to be independent. Definition~\ref{def:bias} clarifies the concepts of unbiased, biased, and fully-biased (S, Z) BIPs/channels.
\begin{mydef} [Unbiased/Biased Channels]
\label{def:bias}
Channel $C_i$ is said to be \textbf{unbiased} if $\alpha_i = \beta_i$, and \textbf{biased} if $\alpha_i \neq \beta_i$. Fully-biased channels are special cases of biased channels: an S-channel with $\beta_i = 0$~(Fig.~\ref{fig:ch_s}); a Z-channel with $\alpha_i = 0$~(Fig.~\ref{fig:ch_z}). 
%
%
%
\begin{figure}[!b]
\centering
\begin{subfigure}{0.115\textwidth}
\includegraphics[width=\textwidth]{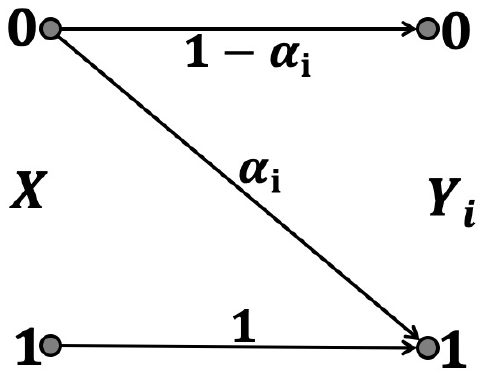}
\caption{S-Channel} \label{fig:ch_s}
\end{subfigure}
\qquad
\begin{subfigure}{0.115\textwidth}
\includegraphics[width=\textwidth]{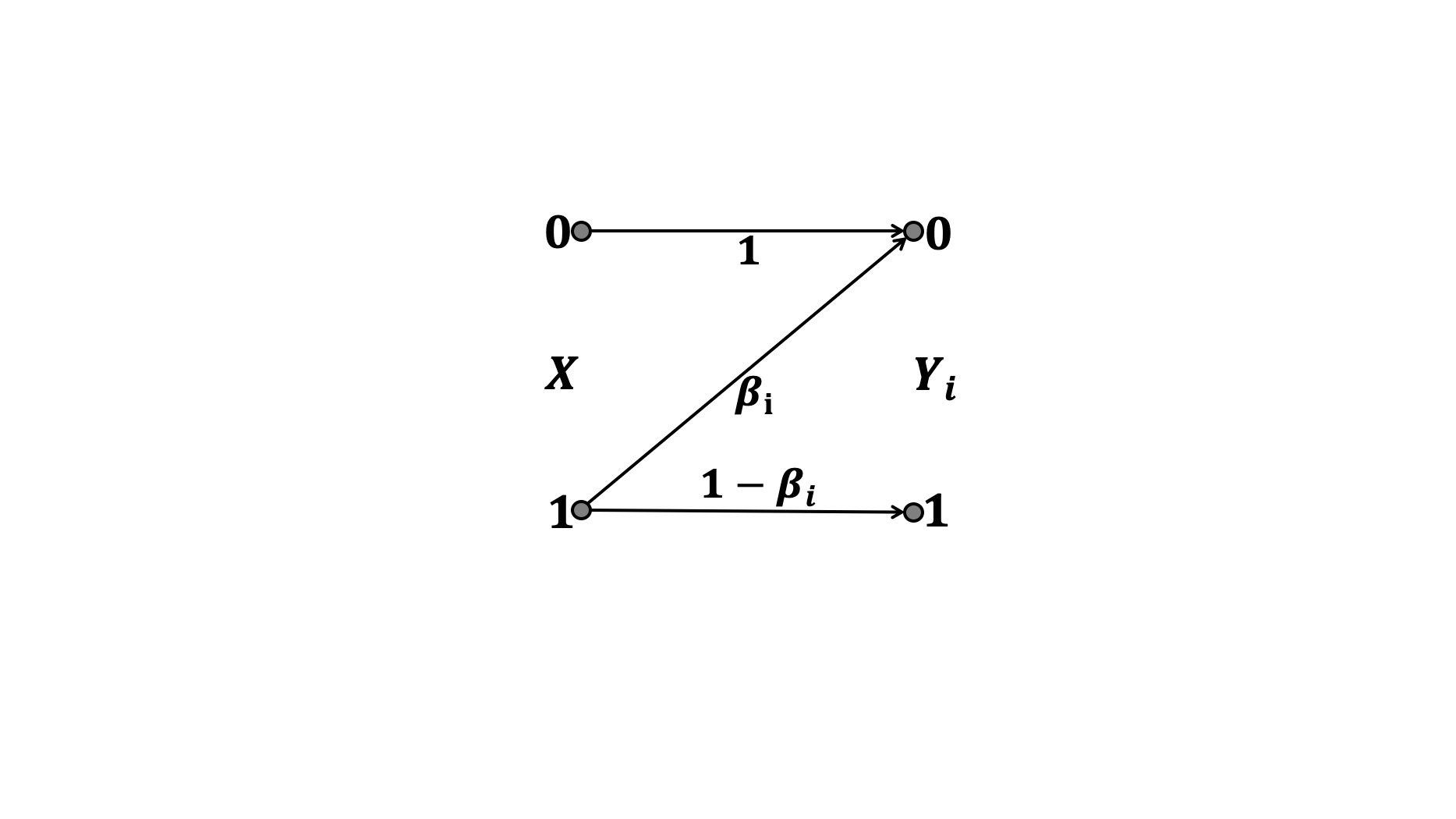}
\caption{Z-Channel} \label{fig:ch_z}
\end{subfigure}
\caption{S- and Z-Channels} \label{fig:ch_sz}
\end{figure}
\end{mydef}
\noindent \textbf{Channel Error Rate:} Given the priors $\rhovect = (\rho_0,\rho_1)$, the average rate at which an erroneous output is received at \textbf{d} from channel $C_i$ is given by
\begin{align}
\begin{array}{rcl}
r_i^{\left( \rhovect \right)} &:=& \rho_0 \alpha_i + \rho_1 \beta_i.
\end{array} \label{eq:rate}
\end{align}
Also, in vector form, we use $\vrn = \{ r_i^{\left( \rhovect \right)} \}_{i = 1}^n$.

%
%
%
In the next section, we will first characterize the optimal decision rule for an RIC receiving information from independent BIPs. Then, we will reveal the impact of the biases $(\van,\vbn)$, the priors $\rhovect$, and the error rates $\vrn$ on the RIC's behavior and the least probability of error it hopes to achieve.
\section{Main Results} \label{sec:main}
%
%
In this section, we will present our main findings on how bias (Defn.~\ref{def:bias}) affects the set of $n$ BIPs that a graph-blind RIC expects to be its optimal choice. This set of $n$ BIPs is modeled by $\vsn$, a system of $n$ parallel, independent binary channels. In particular, after characterizing the optimal decision policy, we will first expand on the counter-intuitive fact that an RIC will choose the most biased BIPs to minimize its probability of consuming false information (cf. Theorem~\ref{thm:concave}). Then, we will discuss the exponentially growing gains that the RIC expects to obtain by choosing fully-biased BIPs over unbiased ones (cf. Theorem~\ref{thm:gains}). The proofs of these results follow in Section~\ref{sec:lowerbound}.
\subsection{Characterization and Discussion of the Decision Rule}
\begin{figure}
\centering
\includegraphics[width=0.28\textwidth,keepaspectratio]{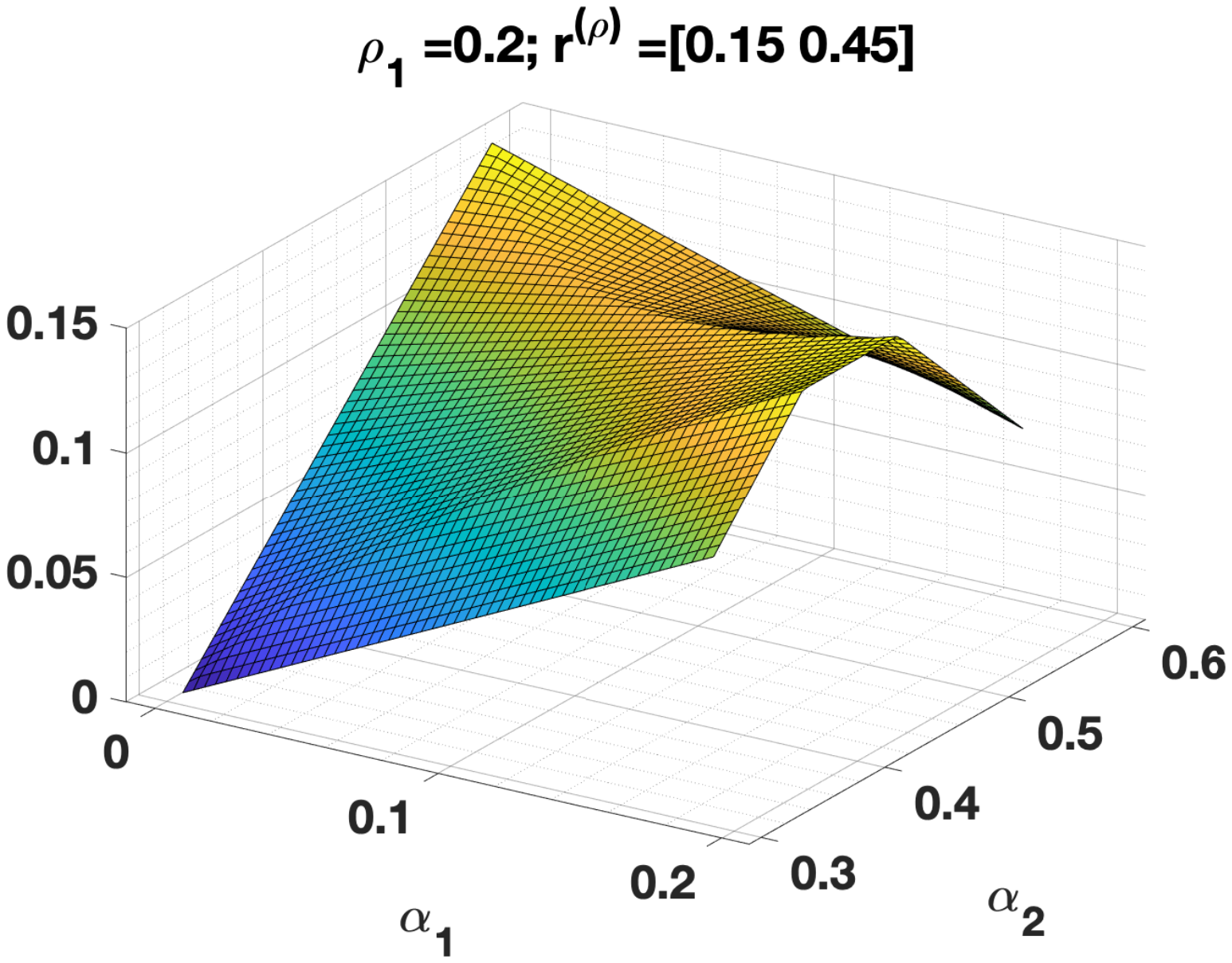}
\caption{Minimum error probability for $2$ independent channels with fixed error rate, but varying bias reveals a complex form.}
\label{fig:errorprobn2}
\end{figure}

%
%
We start by describing the nature of the decision rule that forms the optimal decision policy for $\vsn$. We define 
\begin{align}
&\proda := \proda \left(\vyn\right) := P(\vyn | X = 0) = \overset{n}{\underset{i = 1}{\prod}} \alpha_i^{y_i} \left( 1-\alpha_i \right)^{1-y_i},\notag\\
&\prodb := \prodb \left(\vyn\right) := P(\vyn | X = 1) = \overset{n}{\underset{i = 1}{\prod}} \left( 1-\beta_i \right)^{y_i} \beta_i^{1-y_i}.\notag
\end{align}
The classical hypothesis testing framework \cite{poor2013introduction} yields the optimal decision rule $\pi^{\ast}(\vyn)$ for a given $\vyn$ as follows:
\begin{align}
& \rho_0 \proda \left(\vyn\right) \overset{\pi^{\ast}(\vyn) = 1}{\underset{\pi^{\ast}(\vyn) = 0}{\lessgtr}} \rho_1 \prodb \left(\vyn\right), ~ \forall \vyn.\label{eq:decrule}
\end{align}
This decision rule can be further simplified into a Log-Likelihood-Ratio (LLR) with an additive structure. However, it is easy to see that this test is a highly nonlinear function of the bias parameters: $(\van, \vbn)$. This nonlinearity significantly complicates the RIC's error analysis of the decision rule w.r.t. bias, which is the main objective of this work. 
%
%
To that end, consider the impact of a change in $\alpha_i$ on the terms in \eqref{eq:decrule}: 
%
\begin{align*}
&\frac{\partial \log \left( \proda/\prodb \right) }{\partial \alpha_i} = \frac{(\rho_1 - r_i^{\left( \rhovect \right)})y_i}{\rho_1 \alpha_i (1-\beta_i)} + \frac{(\rho_0 - r_i^{\left( \rhovect \right)})(1-y_i)}{\rho_1 \beta_i (1-\alpha_i)}.
\end{align*}
In particular, note that: The value of $y_i$ \big(not $r_i^{(\rhovect)}$\big) decides if $\proda$, $\prodb$ are both increasing (decreasing) in $\alpha_i$. So, even when $\frac{\proda}{\prodb}$ is monotonic in ${\alpha_i, \: \forall i}$ for all $\vyn$, the impact of $\van$ on $P_e^{\pi^{\ast}} \left( \vsn \right) = \sum_{\vyn} \min \left\{ \proda, \prodb \right\}$ might not be monotonic. Fig.~\ref{fig:errorprobn2} illustrates this non-trivial nature of the RIC's (perceived) probability of error as a function of biases, for $n$ as small as $2$, calling for a comprehensive analysis of the RIC's choices.


\subsection{Impact of Channel Biases on Optimal Performance}

Realizing the difficulty in the direct analysis of the optimal decision rule on $\vsn$, we are motivated to seek a uniform lower bound on an RIC's anticipated error probability. By proving that the RIC's anticipated error probability, from the optimal decision rule, is coordinate-wise concave in $\van$, Theorem~\ref{thm:concave} leads us to the fact that: \emph{an RIC anticipates to maximize its probability of detecting the ground truth, by choosing BIPs that are fully-biased against the a-priori likely outcome.} 




\begin{mythm}[Performance of the optimal decision policy is coordinate-wise concave in bias]~
\label{thm:concave}

Without loss of generality, assume that $\rho_0 \geq \rho_1$. Consider a system of $n$ independent, parallel binary channels described by $\vsn = \left( n,\rho_0,\van,\vbn \right)$, where $\rho_0 \van + \rho_1 \vbn = \vrn$. Assume that $\vrn$ is fixed to be a constant, and\footnote{If $\exists k$ such that $r_k^{(\rhovect)} > \half$, then we can map $(1-Y_k, 1-r_k^{(\rhovect)}, 1-\alpha_k, 1-\beta_k) \mapsto (Y_k, r_k^{(\rhovect)}, \alpha_k, \beta_k)$.}
$r_k^{(\rhovect)} \in \left[ 0, \half \right] \forall k$.

For every such $\vsn$, there is an optimal decision policy $\pi^{\ast}$ chosen from $\Pi^{\ast}  \left( \vsn \right)$ and an average probability of error $P_e^{\pi^{\ast}} (\vsn)$. This error function, $P_e^{\pi^{\ast}} (\vsn)$, is \emph{concave} in $\alpha_k$ for any $k \in \{1,2,\cdots,n\}$, when $\alpha_{i}$ is fixed $\forall \; i \neq k$.

\end{mythm}

Theorem~\ref{thm:concave} leads us to the conclusion that: while holding the $\vrn$ fixed, the error function $P_e^{\pi^{\ast}}(\vsn)$ will achieve its least value for a system $\vsn$ such that $\alpha_k \in \left\{ \alpha_{k,\min}, \; \alpha_{k,\max} \right\}~\forall k$. Here, $\alpha_{k, \min} \geq \frac{\left( r_k^{(\rhovect)} - \rho_1 \right )^+}{\rho_0}$ and $\alpha_{k,\max} \leq \frac{r_k^{(\rhovect)}}{\rho_0}$.
We will refine the $2^n$ possible combinations of extreme bias to find the one that achieves the lower bound on the probability of error.

\begin{mycorr}[Similarly and fully-biased BIPs are optimal]
\label{thm:extremebias}

Assume that $\rho_0 \geq \rho_1$, $\vrn \in \left[0,\half\right]^n$ is fixed, and $\displaystyle \rho_0 \prod_{i = 1}^n \frac{r_i^{(\rhovect)}}{\rho_0} \leq \rho_1$. Then the least probability of error is achieved by an optimal policy on $n$ parallel, binary channels when $\vsn$ is a system of $n$ S-channels (Fig.~\ref{fig:ch_s}). That is, $\vbn = \vzn$.
\begin{equation}
\label{eq:leasterror}
P_e^{\pi^{\ast}}(\vsn) \geq \rho_0 \prod_{i = 1}^n \frac{r_i^{(\rhovect)}}{\rho_0}.
\end{equation}
Since $\rho_0 \geq \rho_1$, the optimal decision policy for a system of S-channels is $\pi^{\ast} \left( \vYn \right) = Y_1 \cdot Y_2 \cdot \cdots \cdot Y_n$.
\end{mycorr}
Corollary~\ref{thm:extremebias} strongly reveals that an RIC prefers systems with $n$ independent BIPs that are fully-biased against the a-priori likely outcome (i.e., BIPs with S or Z-channels, depending on the priors) to all other BIPs with the same individual error rates. \emph{An RIC with such BIPs will assume that it has identified the ground truth without making an error, unless all $n$ BIPs communicate the a-priori unlikely value.}


This result provides an insight that may be counter-intuitive: RICs looking for $n$ independent information-providers with prefixed error rates $\vrn$, find that it is not optimal to choose the diversely-biased or unbiased BIPs! In fact, the RIC finds it is best to select all similarly and fully-biased BIPs, but pay extra attention if one of them reports \textit{against} their bias. 

We note that this finding greatly extends a loosely related result in \cite{silverman1955binary}, which proves that only among channels of very low and equal capacity, a maximally asymmetric channel with a noiseless symbol has the least probability of error.
\subsection{Gains of Fully-Biased Channels over Unbiased Channels}
\label{sec:gains}
Motivated by the optimality of full bias when BIPs are independent, we also studied the gains the RIC anticipates after replacing unbiased BIPs with a set of similarly and fully-biased BIPs. For fair comparison, we assume that the average error rates $r_i^{\left( \rhovect \right)}$ of the BIPs are all equal to $r^{\left( \rhovect \right)}$ in both scenarios. Therefore, the BIPs are equivalent in their average rate of sending erroneous bits, but differ in their biases. 

Theorem~\ref{thm:gains} obtains the upper and lower bounds on the RIC's anticipated gains, and shows that these bounds become 
tight as $n$ (the number of BIPs sought by the RIC) increases. 

\begin{mythm}[Gains of Fully-Biased BIPs vs Unbiased BIPs]
\label{thm:gains}

Fix $r^{\left( \rhovect \right)} \in \left( 0, \half \right]$ to be the common error rate for all the BIPs' channels. Then, let $\vsn^u= \left\{ n, \rho_0, r^{\left( \rhovect \right)} \vin, r^{\left( \rhovect \right)} \vin \right\}$ and $\vsn^f= \left\{ n, \rho_0, \frac{r^{\left( \rhovect \right)}}{\rho_0} \vin, \vzn \right\}$, respectively, describe the \textit{unbiased} and \textit{fully-biased} systems, each containing a set of $n$ independent BIPs. Correspondingly, let the error-optimal decision policies for $\vsn^u$ and $\vsn^f$ be denoted as $\pi^{\ast}_u$ and $\pi^{\ast}_f$, respectively\footnote{Note that $\pi^{\ast}_f$ is the policy described in Corollary~\ref{thm:extremebias}.}. Then, for any $n\in \nat$, we have
\begin{align*}
& \ln\left(\frac{P_e^{\pi^{\ast}_u} \left( \vsn^u \right)}{P_e^{\pi^{\ast}_f} \left( \vsn^f \right)}\right) \leq  m \ln \left( 4\rho_0^2\left(\frac{1}{r^{\left( \rhovect \right)}}-1\right) \right) + \ln \frac{2(m+1)}{\rho_0}
\end{align*}
\begin{align*}
&\ln\left(\frac{P_e^{\pi^{\ast}_u} \left( \vsn^u \right)}{P_e^{\pi^{\ast}_f} \left( \vsn^f \right)}\right) \geq m \ln \left( 4 \rho_0^2 \left(\frac{1}{r^{\left( \rhovect \right)}} - 1 \right) \right) - \ln \frac{4m}{\rho_1},
\end{align*}
where $m = \modfloor{\frac{n}{2}}$. Moreover, asymptotically, these bounds converge to get
\begin{align}
\displaystyle \lim_{n\rightarrow \infty} \frac{1}{n}\ln \left( \frac{P_e^{\pi^{\ast}_u} \left( \vsn^u \right)}{P_e^{\pi^{\ast}_f} \left( \vsn^f \right)} \right) = \frac{1}{2} \ln \left( 4 \rho_0^2 \left(\frac{1}{r^{\left( \rhovect \right)}} - 1 \right) \right).\label{eq:tight}
\end{align}
\end{mythm}

This theorem reveals that the RIC expects the proportional gains, of using $n$ fully-biased BIPs rather than $n$ unbiased BIPs, to increase exponentially with $n$ and the factor that is \textit{explicitly} characterized in \eqref{eq:tight} (in terms of the average error rates $r^{\left( \rhovect \right)}$ of the channels). This shows that the RIC's anticipated gains are particularly high when the average error rates $r^{\left( \rhovect \right)}$ are closer to zero. It is expected that the (anticipated) gains will be relatively small when $r \to \half$, since the probability of error approaches $\half$ in both scenarios.

This result is quite useful in characterizing the conditions under which the RIC would prefer to use fully-biased BIPs as opposed to unbiased ones, and the conditions under which using unbiased BIPs may be acceptable to the RIC. Moreover, the upper and lower bounds on the RIC's anticipated gains can be reverse engineered to determine the number of BIPs $n$ that the RIC will choose to guarantee a desired limit on the probability of error.

\section{Lower Bound on Error Probability}
\label{sec:lowerbound}

In this section, our objective is to fix the channel error rates $\vrn$ and select the biases $(\van, \vbn)$ which minimize the error probability achieved by an optimal decision policy $\pi^{\ast} \in \Pi^{\ast} \left( \vsn \right)$.
%
%
%
For ease of presentation in the proofs, we define:
\begin{align*}
\prodab := \prodab \left( \vYn \right) &:= \proda \left( \vYn \right) - \prodb \left( \vYn \right), \\
\proddiv := \proddiv \left( \vYn \right) &:= {\proda \left( \vYn \right)} \slash {\prodb \left( \vYn \right)}.
\end{align*}
We also identify sets of outcomes, $\vYn$, which elicit identical decisions from a decision policy $\pi$. Here, $\pi$ need not be an optimal decision policy for a given ${\vsn = \left( n,\rho_0, \van, \vbn \right)}$.
\begin{align*}
& \ct_a^{\pi} := \left\{ \vYn \in \{0,1\}^n \mid \pi \left( \vYn \right) = a \right\}, \\
& \ct_{a,Y_k=b}^{\pi} := \left\{ \vYnk \in \{0,1\}^{n-1} \mid \pi \left( \vYnk, b \right) = a \right\}, 
\end{align*}
where ${a, b \in \{0,1\}}$, ${\vYn_{-k} = \left[ Y_1, \cdots, Y_{k-1}, Y_{k+1}, \cdots, Y_n \right]^T}$.

\subsection{Coordinate-wise concavity of the probability of error} 
Here, we will prove that when the individual error rates are prefixed at $\vrn$ and the biases are varying, the probability of error, $P_e^{\pi^{\ast}}(\vsn)$,  is coordinate-wise concave in each BIP's bias.
\begin{proof}[\emph{\underline{Proof of Theorem~\ref{thm:concave}}}]

The probability of error achieved by an optimal decision policy $\pi^{\ast} \in \Pi^{\ast}  \left( \vsn \right)$ in~\eqref{eq:minerrprob} is restated as:
\begin{align*}
P_e^{\pi^{\ast}}(\vsn) ~ &= ~ \underset{\cta}{\sum} \rho_1 \prodb (\vYn) + \underset{\ctb = \left( \cta \right)^C}{\sum} \rho_0 \proda (\vYn)\\
&= ~ \rho_1 + \underset{\ctb}{\sum} \left( \rho_0 \proda (\vYn) - \rho_1 \prodb (\vYn) \right).
\end{align*}

For a decision policy $\pi$ to be an optimal on $\vsn$, we need: 
\begin{align}
\proddiv < \rho_1 / \rho_0 \implies \vYn \in \ct_1^{\pi} \implies \proddiv \leq \rho_1 / \rho_0.\label{eq:optdecpolicy}
\end{align}


In order to prove this theorem, we will proceed by fixing the values of $k$ and $\van_{-k}$ and then obtaining the partial derivative  $P_e^{\pi^{\ast}}(\vsn)$ with respect to $\alpha_k$ for some arbitrary $k$.

The function $P_e^{\pi^{\ast}} (\vsn)$ is {continuous and piecewise-linear}, but non-differentiable at those values of $\alpha_k$ for which ${\exists\vYn:\proddiv \left(\vYn\right)=\rho_1 / \rho_0}$. 
So, for a given $k$ and fixed $\van_{-k}$, consider some interval 
\begin{align}
&\alpha_k \in \left[ \alpha_k^L, \alpha_k^R \right] \text{ s.t. } \left\{\vYn:\proddiv = \frac{\rho_1}{\rho_0} \right\} = \phi. \label{eq:alpha_intervals}
\end{align}
So, in each such interval, $P_e^{\pi^{\ast}} (\vsn)$ is always differentiable and the optimal policy $\pi^{\ast}$ is constant. In such an interval, the partial derivative of $P_e^{\pi^{\ast}}(\vsn)$ w.r.t. $\alpha_k$ is: 
\begin{align*}
\frac{\partial P_e^{\pi^{\ast}}(\vsn)}{\rho_0 \cdot \partial \alpha_k} &= \underset{\ctbk{1}}{\sum} \prodab \left( \vYnk \right) - \underset{\ctbk{0}}{\sum} \prodab \left( \vYnk \right).
\end{align*}
Here, $\frac{\partial \beta_k}{\partial \alpha_k} = -\frac{\rho_0}{\rho_1}$ since $\rho_0 \alpha_k + \rho_1 \beta_k$ is a constant.

Now, it remains for us to determine whether the partial derivative $\frac{\partial P_e^{\pi^{\ast}}(\vsn)}{\rho_0 \cdot \partial \alpha_k}$ is a monotonic non-increasing function on $\alpha_k$, regardless of our choice of $k$ and $\van_{-k}$.

The partial derivative depends on the sets $\ctbk{0}$, $\ctbk{1}$ and the value of $\prodab$ on these sets. The subsets of outcomes, $\ctbk{1}$ and $\ctbk{0}$, relate to each other as follows: 
\begin{align}
\ctbk{0} \underset{\supset}{\subseteq} \ctbk{1} & \iff \alpha_k + \beta_k \lesseqgtr 1.\label{eq:Tab}\\
\text{Since, } \alpha_k + \beta_k \lesseqgtr 1 & \iff \frac{\rho_1 \beta_k}{\rho_0 (1-\alpha_k)} \lesseqgtr \frac{\rho_1 (1-\beta_k)}{\rho_0 \alpha_k}.\label{eq:alphabeta1}
\end{align}
Given~\eqref{eq:optdecpolicy},~\eqref{eq:Tab},~and~\eqref{eq:alphabeta1}, we can now proceed as follows.

\noindent \paragraph{{{WHEN $r_k^{(\rhovect)} \leq \rho_1$}}}~
The condition $\alpha_k + \beta_k \leq 1$ always holds when $r_k^{(\rhovect)} \leq \rho_1$. Using~\eqref{eq:Tab}, 
\begin{equation}
\label{eq:rlessrho1}
\frac{\partial P_e^{\pi^{\ast}}(\vsn)}{\rho_0 \cdot \partial \alpha_k} = \underset{\ctbk{1} \setminus \ctbk{0}}{\sum} \prodab \left( \vYnk \right).
\end{equation}
%

In addition, for a channel $\vc_k$ with $r_k^{(\rhovect)} \leq \rho_1$, the following is true regardless of the bias $(\alpha_k,\beta_k)$:

${\left( 1 + \frac{\rho_0 - r_k^{(\rhovect)}}{\rho_1 \beta_k} \right)^{-1} = \frac{\rho_1 \beta_k}{\rho_0 (1-\alpha_k)} \leq 1 \leq \frac{\rho_1 (1-\beta_k)}{\rho_0 \alpha_k} = 1 + \frac{\rho_1 - r_k^{(\rhovect)}}{\rho_0 \alpha_k}.}$

Moreover, in this case, if $\vYnk \in \ctbk{1} \setminus \ctbk{0}$ then:

\hspace{0.4in} $\displaystyle \frac{\rho_1 \beta_k}{\rho_0 (1-\alpha_k)} \leq \proddiv \left( \vYn \right) \leq \frac{\rho_1 (1-\beta_k)}{\rho_0 \alpha_k}.$

\begin{enumerate}

\item[(i)] $\frac{\rho_1 (1-\beta_k)}{\rho_0 \alpha_k} - 1 = \frac{\rho_1 - r_k^{(\rhovect)}}{\rho_0 \alpha_k} \geq 0$ is decreasing with $\alpha_k$, 

\item[(ii)] $1 - \frac{\rho_1 \beta_k}{\rho_0 (1-\alpha_k)} = \frac{\rho_0 - r_k^{(\rhovect)}}{\rho_0 (1-\alpha_k)} \geq 0$ is increasing with $\alpha_k$.

\end{enumerate}

Thus, in~\eqref{eq:rlessrho1}, the set $\ctbk{1} \setminus \ctbk{0}$ will \emph{accumulate} terms for which $\prodab \left( \vYnk \right)$ is negative and \emph{discard} terms for which $\prodab \left( \vYnk \right)$ is positive, as the values of $\alpha_k^L$ and $\alpha_k^R$ that create the interval $\alpha_k$ in~\eqref{eq:alpha_intervals} increase.

Therefore, $P_e^{\pi^{\ast}}(\vsn)$ is concave in $\alpha_k$, when $r_k^{(\rhovect)} \leq \rho_1$.

\paragraph{{{WHEN $r_k^{(\rhovect)} \geq \rho_1$}}} ~
Notice that the sum $1 + \frac{\partial \beta_k}{\partial \alpha_k} = 1 - \frac{\rho_0}{\rho_1}$ is non-positive. Consequently, $\alpha_k + \beta_k$ is a non-increasing function of $\alpha_k$.

Consequently, ${\alpha_k + \beta_k \geq 1}$ if and only if $\alpha_k \in \left[\frac{(r_k^{(\rhovect)} - \rho_1)^+}{\rho_0},  \frac{r_k^{(\rhovect)} - \rho_1}{\rho_0 - \rho_1} \right]$. Applying~\eqref{eq:Tab}, we get (a) and (b).

\begin{enumerate}

\item[(a)] In the interval $\alpha_k \in \left[\frac{(r_k^{(\rhovect)} - \rho_1)^+}{\rho_0},  \frac{r_k^{(\rhovect)} - \rho_1}{\rho_0 - \rho_1} \right]$, we get: 
\begin{equation}
\label{eq:rgreatrho1sumgreat1}
\frac{\partial P_e^{\pi^{\ast}}(\vsn)}{\rho_0 \cdot \partial \alpha_k} = \underset{\ctbk{0} \setminus \ctbk{1}}{\sum} -\prodab \left( \vYnk \right).
\end{equation}

\item[(b)] Using~\eqref{eq:Tab} in the interval $\alpha_k \in \left[ \frac{r_k^{(\rhovect)} - \rho_1}{\rho_0 - \rho_1}, \frac{r_k^{(\rhovect)}}{\rho_0} \right]$, we get that~\eqref{eq:rlessrho1} holds true for these $\alpha_k$.

\end{enumerate}


Moreover, in this case, whether ${\vYnk \in \ctbk{1} \setminus \ctbk{0}}$, or ${\vYnk \in \ctbk{0} \setminus \ctbk{1}}$ we have:

\hspace{0.25in} $\displaystyle \proddiv \left( \vYn \right) \leq 1 \implies \prodab \left( \vYn \right) \leq 0.$

\begin{enumerate}

\item[(i)] $1 - \frac{\rho_1(1-\beta_k)}{\rho_0 \alpha_k} = \frac{ r_k^{(\rhovect)} - \rho_1}{\rho_0 \alpha_k} \; \geq 0$ is decreasing with $\alpha_k$,  
\item[(ii)] $1 - \frac{\rho_1 \beta_k}{\rho_0 (1-\alpha_k)} = \frac{\rho_0 - r_k^{(\rhovect)}}{\rho_0(1-\alpha_k)} \; \geq 0$ is increasing with $\alpha_k$.

\end{enumerate}



Thus, when $\alpha_k \leq \frac{r_k^{(\rhovect)} - \rho_1}{\rho_0 - \rho_1}$ and~\eqref{eq:rgreatrho1sumgreat1} holds true:
The set $\ctbk{0} \setminus \ctbk{1}$ will \emph{discard} more terms for which $\prodab \left( \vYnk \right)$ is negative and $\frac{\partial P_e^{\pi^{\ast}}(\vsn)}{\partial \alpha_k}$ will become less positive, as the values of $\alpha_k^L$ and $\alpha_k^R$ that create the interval $\alpha_k$ in~\eqref{eq:alpha_intervals} increase.




And, when $\alpha_k \geq \frac{r_k^{(\rhovect)} - \rho_1}{\rho_0 - \rho_1}$ and~\eqref{eq:rlessrho1} holds true:
The set $\ctbk{1} \setminus \ctbk{0}$ will \emph{accumulate} more terms for which $\prodab \left( \vYnk \right)$ is negative and $\frac{\partial P_e^{\pi^{\ast}}(\vsn)}{\partial \alpha_k}$ will become more negative, as the values of $\alpha_k^L$ and $\alpha_k^R$ that create the interval $\alpha_k$ in~\eqref{eq:alpha_intervals} increase.

Therefore, $P_e^{\pi^{\ast}}(\vsn)$ is concave in $\alpha_k$, when $r_k^{(\rhovect)} \geq \rho_1$.


Thus, assuming $\vrn \in \left[ 0, \half \right]^n$ and it is fixed, the error function $P_e^{\pi^{\ast}}(\vsn)$ is coordinate-wise concave w.r.t. $\alpha_k$.
\end{proof}
\underline{Note}: Theorem~\ref{thm:concave} does NOT prove that the error function, $P_e^{\pi^{\ast}}(\vsn)$, is jointly concave with respect to $\van$.

\underline{Note}: It is obvious that the value $\alpha_k = \frac{r_k^{(\rhovect)} - \rho_1}{\rho_0 - \rho_1}$ represents the local maxima of the error function, whenever $r_k^{(\rhovect)} \geq \rho_1$ regardless of the choice of $\vrn_{-k}$ and $\van_{-k}$.

Theorem~\ref{thm:concave} leads us to the conclusion that: 
assuming $\rho_0 \geq \rho_1$ and $\vrn$ is fixed, the probability of error $P_e^{\pi^{\ast}}(\vsn)$ is at its minimum for a system $\vsn$, where $\alpha_k \in \left\{ \alpha_{k,\min}, \alpha_{k,\max} \right\}, ~ \forall k$. Here, $\alpha_{k, \min} \geq \frac{\left( r_k^{(\rhovect)} - \rho_1 \right )^+}{\rho_0}$ and $\alpha_{k,\max} \leq \frac{r_k^{(\rhovect)}}{\rho_0}$.

%
%
\subsection{Optimality of similarly-biased BIPs holding maximum bias}
Now, we will seek from the $2^n$ possible combinations of the most extreme biases, the combination that will represent the lower bound on the probability of error to the RIC.

\begin{proof}[\emph{\underline{Proof of Corollary~\ref{thm:extremebias}}}]

From Theorem~\ref{thm:concave}, we know that the error function $P_e^{\pi^{\ast}}(\vsn)$ is concave in each $\alpha_k$ for every $\van_{-k}$. Therefore, for a fixed $\vrn$, we can conclude that the lower bound on the error function $P_e^{\pi^{\ast}}(\vsn)$ can only occur at an extreme bias. That is, when $\alpha_k \in \left\{ \frac{\left( r_k^{(\rhovect)} - \rho_1 \right )^+}{\rho_0}, \frac{r_k^{(\rhovect)}}{\rho_0} \right\}, ~ \forall k$. For each of these possible extreme values of $\alpha_k$, at least one value among $\alpha_k$, $\beta_k$, and $1-\beta_k$ is zero.


Thus, whenever an extreme bias is chosen for the binary channel $\vc_i$, only one value of the outcome $Y_i$ can be generated by both $X = 0$ and $X = 1$ while the other is generated by either $X = 0$ or $X = 1$.
So, when we choose extreme biases for every BIP in $\vsn$, the optimal decision policy is prone to error only for one particular $\vYn = \vyn$. For the channel $\vc_i$:
%
\begin{align*}
P(y_{i} \mid X = 0) &= \left\{ \begin{array}{l}
\left( \frac{1-r_i^{(\rhovect)}}{\rho_0} \right)^{1-y_{i}} \left( \frac{r_i^{(\rhovect)}}{\rho_0} \right)^{y_{i}}, \text{ if } r_i^{(\rhovect)} > \rho_1\\
\left( 1 \right)^{1-y_{i}} \left( \frac{r_i^{(\rhovect)}}{\rho_0} \right)^{y_{i}}, \text{ if } r_i^{(\rhovect)} \leq \rho_1.
\end{array} \right.\\
P(y_{i} \mid X = 1) &= \left\{ \begin{array}{rl}
\left( 1 \right)^{1-y_{i}} \left( 1 \right)^{y_{i}}, & \text{ if } r_i^{(\rhovect)} > \rho_1\\
\left( \frac{r_i^{(\rhovect)}}{\rho_1} \right)^{1-y_{i}} \left( 1 \right)^{y_{i}}, & \text{ if } r_i^{(\rhovect)} \leq \rho_1.
\end{array} \right.
\end{align*}

Since $r_i^{(\rhovect)} \leq \half$, we always have $1-r_i^{(\rhovect)} \geq r_i^{(\rhovect)}$. Moreover, when $r_i^{(\rhovect)} > \rho_1$, we have $r_i^{(\rhovect)}, 1-r_i^{(\rhovect)} < \rho_0$. Finally, since $\rho_0 \geq \rho_1$, we deduce that the combination of extreme biases that will represent the lower bound of the error function, has an optimal decision policy is error-prone only when $\vYn = \vin$.
%
%
\begin{align*}
\therefore P_e^{\pi^{\ast}}(\vsn) &= \min \left\{ \rho_0 \overset{n}{\underset{k=1}{\prod}} \frac{r_k^{(\rhovect)}}{\rho_0}, ~~ \rho_1 \cdot 1 \right\}.\qedhere
\end{align*}
\end{proof}

Through the results presented in this section, we conclude that: if an RIC wishes to receive an information bit $X$  through \emph{$n$ (presumably, independent) BIPs}, then for prefixed individual error rates, the RIC is best served by $n$ fully-biased BIPs which favor the a-priori unlikely outcome (BIPs are either S or Z type, depending on the priors). Then the RIC is prone to making an error in its deduction only when all $n$ BIPs are wrong. For example, when no sensor sends correct information to the central node; when all the friends and trusted news sources of a person choose to spread false information.

\subsection{Characterizing the gains of fully-, similarly-biased BIPs} 
\label{sec:errmaxpolicy}

Corollary~\ref{thm:extremebias} demonstrates ``how" the BIPs's affect the RIC's decisions. In this section, we will investigate the gains that the RIC anticipates to obtain from its choices.
%
%
%
%
%
\begin{proof}[\emph{\underline{Proof of Theorem~\ref{thm:gains}}}]

We will start with a useful claim. Its proof is omitted due to its simplicity and space constraints.
\begin{myclaim} \label{claim:comb}
$\displaystyle {{2m + 1 \choose m} < 2 {2m \choose m}};~ {{2m \choose m} = 4^m \prod_{j = 1}^m \left( 1 - \frac{1}{2j} \right)}.$
\end{myclaim}
%
%


We also denote $\pi^{\ast}_u (\vyn)$ as $\pi^{\ast}_u (k)$ since the channels are identical \Big(let $k$ be the number of 1's in $\vyn$\Big).
\begin{align}
&P_e^{\pi^{\ast}_u} \left( \vsn^u \right) = \overset{n}{\underset{k = 0}{\sum}} {n \choose k} \min \left\{ \rho_0 (r^{\left( \rhovect \right)})^k (1-r^{\left( \rhovect \right)})^{n-k},\right.\notag\\
& \hspace{1in} \left. \rho_1 (1-r^{\left( \rhovect \right)})^k (r^{\left( \rhovect \right)})^{n-k} \right\}\notag
\end{align}
We define $c := \frac{1}{r^{\left( \rhovect \right)}} - 1$ and $m := \modfloor{\frac{n}{2}}$.
\begin{align*}
P_e^{\pi^{\ast}_u} \left( \vsn^u \right) &= \left( r^{( \rhovect )} \right)^n \overset{n}{\underset{k = 0}{\sum}} {n \choose k} \min \left\{ \rho_0 c^{n-k}, \rho_1 c^k \right\}\\
&= \left( r^{( \rhovect )} \right)^n \overset{n}{\underset{k = 0}{\sum}} {n \choose n-k} \min \left\{ \rho_0 c^k, \rho_1 c^{n-k} \right\}\\
&\leq \half \left( r^{( \rhovect )} \right)^n \overset{n}{\underset{k = 0}{\sum}} {n \choose k} c^k\\
&= \left( r^{\left( \rhovect \right)} \right)^n \left[ \sum_{k = 0}^m {n \choose k} c^k
- \rho_0 \indicator_{\{n = 2m\}} {n \choose m} c^m \right].
\end{align*}

From Corollary~\ref{thm:extremebias}, we know the minimum probability of error for $\vsn^f$ with all S-channels is:   $P_e^{\pi^{\ast}_f} \left( \vsn^f \right) = \rho_0^{1-n} (r^{\left( \rhovect \right)})^n$. 
\begin{align}
& \therefore \frac{P_e^{\pi^{\ast}_u} \left( \vsn^u \right)}{P_e^{\pi^{\ast}_f} \left( \vsn^f \right)} \leq 
\rho_0^{n-1} \left[ \sum_{k = 0}^{m} {n \choose k} c^k
- \rho_0 \indicator_{\{n = 2m\}} {n \choose m} c^m \right].\notag
\end{align}

\noindent \textbf{\emph{For the upper bound:}} Using Claim~\ref{claim:comb}, 
\begin{align*}
&\frac{P_e^{\pi^{\ast}_u} \left( \vsn^u \right)}{P_e^{\pi^{\ast}_f} \left( \vsn^f \right)} \leq \rho_0^{n-1} \sum_{k = 0}^{m} {n \choose k} c^k \leq \rho_0^{n-1} (m+1) {2m+1 \choose m} c^m\\
& < \rho_0^{2m-1} (m+1) 2 {2m \choose m} c^m\\
&= \exp \left( m \ln \left( 4 \rho_0^2 c \right) + \ln(m+1) + \ln \left( \frac{2}{\rho_0} \right) \right). 
\end{align*}
%
%
%
%

\noindent \textbf{\emph{For the lower bound:}} Define $p := \modceil{\log_2 m}$.
\begin{align*}
& \frac{P_e^{\pi^{\ast}_u} \left( \vsn^u \right)}{P_e^{\pi^{\ast}_f} \left( \vsn^f \right)} ~ \geq ~ \rho_0^{1-n} \sum_{k = 0}^{m} {n \choose k} \rho_1 c^k \geq \smallratio \rho_0^{2m+1} {2m \choose m} c^m \\
& \geq \rho_1 \left( \rho_0^2 c \right)^m 4^m \prod_{j = 1}^m \left( 1 - \frac{1}{2j} \right) \: \because \text{ (Claim~\ref{claim:comb})}\\
& \geq \rho_1 \left( 4 \rho_0^2 c \right)^m \prod_{i = 1}^p \prod_{j = 2^{i-1}}^{2^i - 1} \left( 1 - \frac{1}{2j} \right) \cdot \left( 1 - 2^{-p-1} \right)\\
& \geq \rho_1 \left( 4 \rho_0^2 c \right)^m \prod_{i = 1}^p \left( 1 - 2^{-i} \right)^{2^{i-1}} \cdot \left( 1 - 2^{-p-1} \right)\\
& \stackrel{(a)}{\geq} \rho_1 \left( 4 \rho_0^2 c \right)^m \left( \half \right)^{p+1} \geq \rho_1 \left( 4 \rho_0^2 c \right)^m e^{-\ln(4m)} \\ 
& = \exp \left(m \ln \left( 4 \rho_0^2 c \right) + \ln (\rho_1) - \ln(4m) \right),
\end{align*}
Inequality $(a)$ is justified since $f(x) := x \ln(1-x/2)$ is non-increasing for $x \in [0,1]$. Thus, $\left( 1-x/2 \right)^x \geq e^{f(1)} = \half$.
\end{proof}
%
%
%
%
%

We note that the lower bound applies for all $n \in \nat$ but its exponent might not be positive for very small $n$. The exponent becomes positive when $n$ is large enough for the fast-growing $m \ln \left( 4 \rho_0^2 c \right)$ term to dominate the slow-growing $\ln(2m) - \ln (\rho_1)$ term.
Further, we observe that as $n$ grows, the \emph{exponents} \emph{of both bounds} \emph{converge} \emph{asymptotically} \emph{to the growth rate} given by $\displaystyle \frac{n}{2} \ln \left( 4 \rho_0^2 \left( \frac{1}{r^{\left( \rhovect \right)}} - 1 \right) \right)$, proving \eqref{eq:tight}.

\section{Simulations} \label{sec:exp}

In this section, we perform numerical experiments to validate our theoretical results and to develop a broader understanding of the impact of bias on the choices that appear optimal to a graph-blind RIC that seeks $n$ BIPs.

We showed that a graph-blind RIC expects to minimize its error probability, by choosing fully-biased BIPs amongst all BIPs that have the same individual error rates (cf. Corollary~\ref{thm:extremebias}).
Further, we want to know if the RIC expects other choices to 
yield vastly different error probabilities (as in Theorem~\ref{thm:gains}). 

\begin{figure}[h!]
\centering
\includegraphics[width=0.35\textwidth]{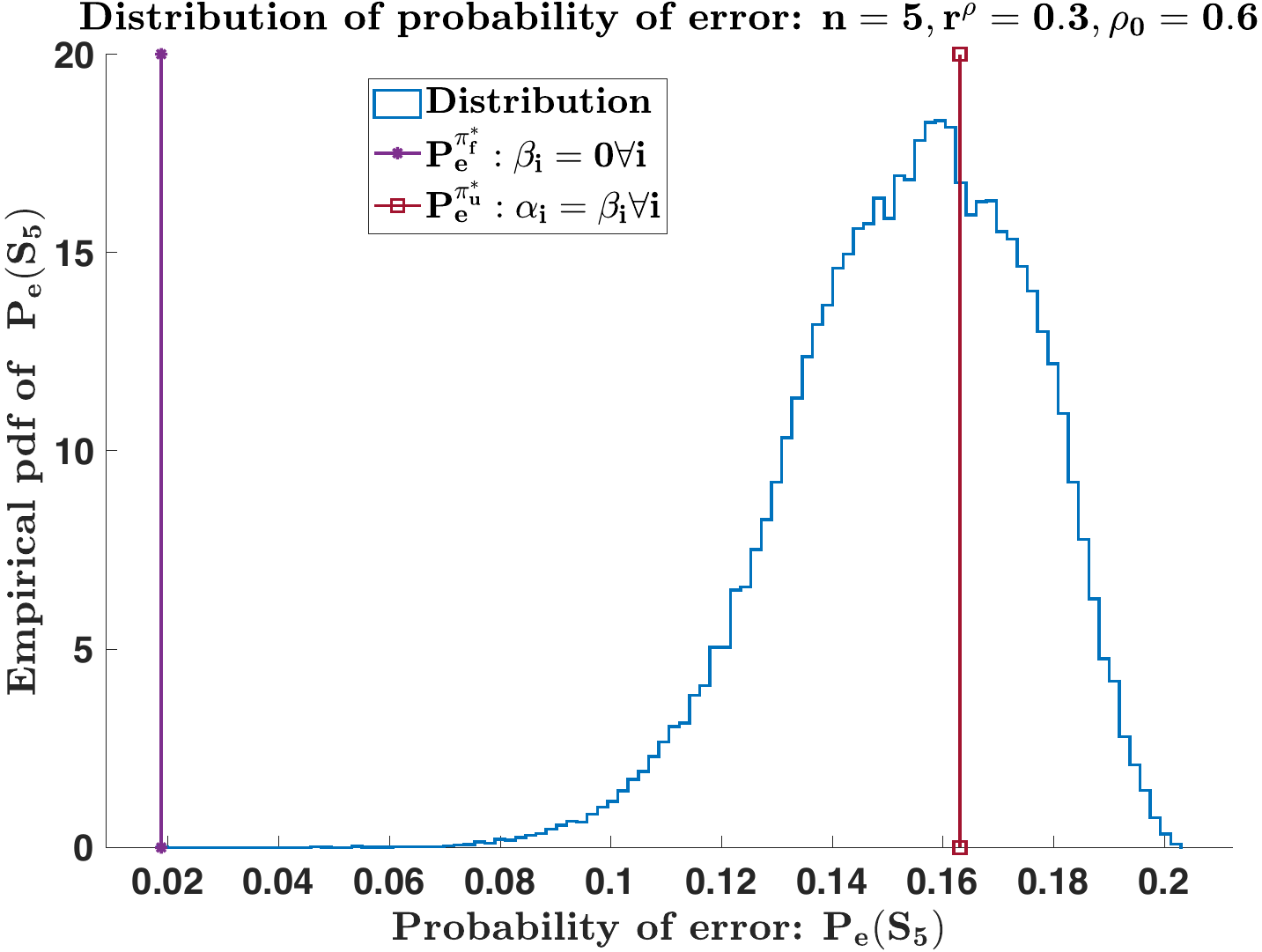}
\caption{Distribution of probabilities of error in a system with fixed priors ($\rhovect$) and fixed channel error rate $r^{\left( \rhovect \right)}$.}
\label{fig:pdf_fullbias_unbias}
\end{figure}

In Figure~\ref{fig:pdf_fullbias_unbias}, we explore this question for $n=5$ channels, $\rhovect = (0.6, 0.4),$ and $r^{\left( \rhovect \right)}=0.3$. We choose many different values of ($\van$, $\vbn$) that satisfy $r^{\left( \rhovect \right)}=0.3$ in order to generate the plotted histogram of the optimal-probability-of-error values obtained by them. The resulting plot confirms that the fully-biased system with all S-channels ($\beta_i=0, \forall i$) achieves the minimum value (in this case $\approx 0.02$). Moreover, it reveals that the error probabilities has a wide range between $0.02$ and $0.2$ with  most of the distribution centered around higher values of $\approx 0.15$. In particular, the case of unbiased channels ($\alpha_i=\beta_i, \forall i$) yields an error probability of $\approx 0.165$. As such, this figure further highlights the significance of utilizing fully-biased channels as opposed to other choices. 

\begin{figure}
\centering
\includegraphics[width=0.36\textwidth]{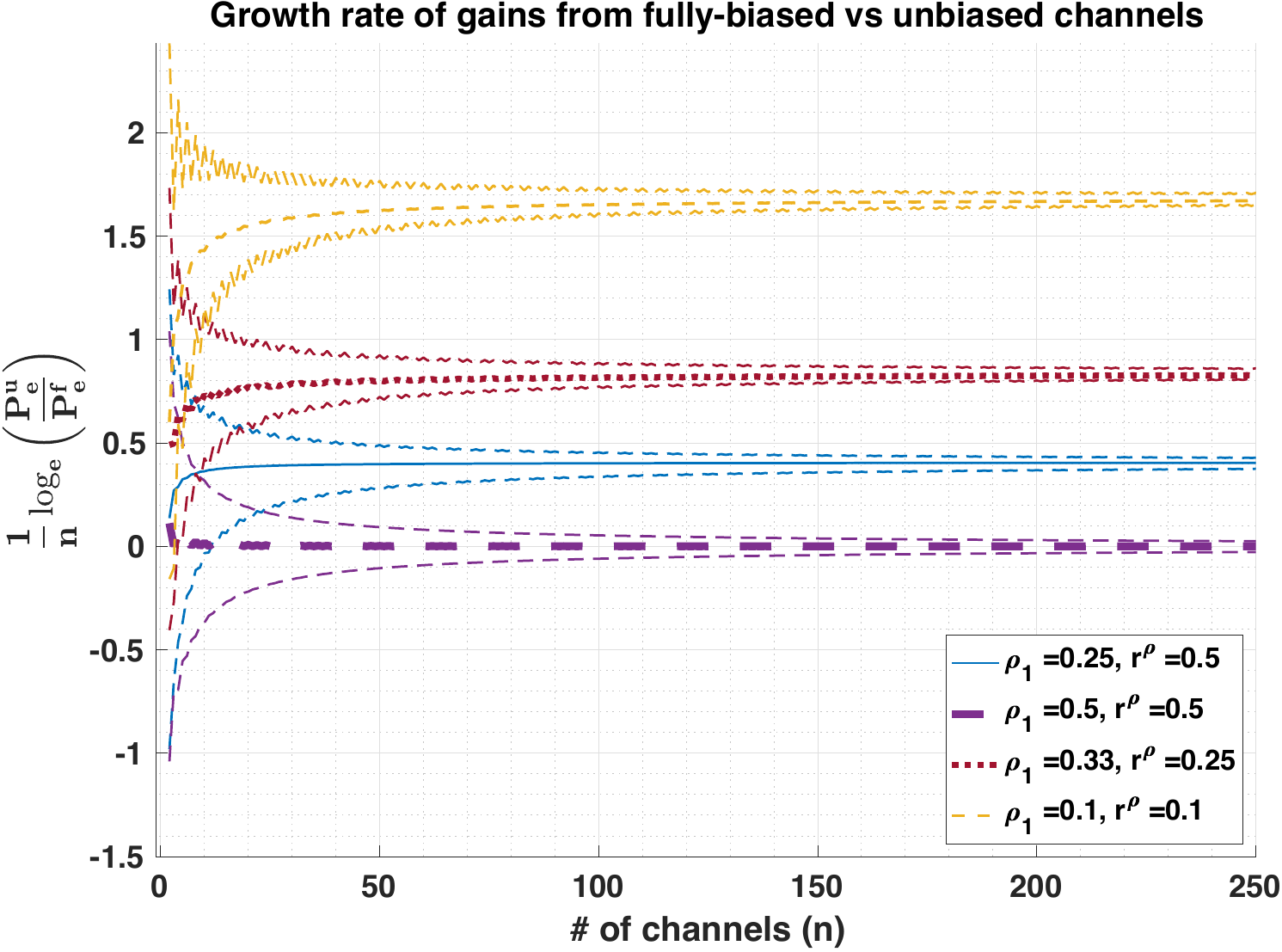}
\caption{Asymptotic growth rate of the gain $\frac{P_e^u}{P_e^f}$.}
\label{fig:gains_bias}
\end{figure}

In Theorem~\ref{thm:gains}, we replaced unbiased BIPs ($\alpha_i = \beta_i$) with fully-biased BIPs that favor the same outcome ($\alpha_i = 0, \forall i$ or $\beta_i = 0, \forall i$). There, we established that the RIC's anticipated gains (factor by which probability of error will fall) grow asymptotically at an exponential rate with the exponent being linearly dependent on $n$ (the number of BIPs/channels used). We obtained this result by deriving upper and lower bounds on the gains that become tight as $n$ increases. To check the tightness of these bounds, the rates of change of the exponents with $n$ are plotted in Figure~\ref{fig:gains_bias} for different $\rhovect$ and $r^{\left( \rhovect \right)}$.
From this plot, we verify that the growth rates of the exponents of the gains are, in fact, linear in $n$. Moreover, the upper and lower bounds are tight even for fairly small values of $n$. We also confirm that the growth rates of the exponents in the gains and their bounds all converge to $\half \ln \left( 4\rho_0^2 \left( \frac{1}{r^{\left( \rhovect \right)}} -1 \right) \right)$. 
Incidentally, the plot also reveals that the RIC expects to obtain almost no gains, by replacing unbiased BIPs with fully-biased BIPs when $\rho_1 = r^{\left( \rhovect \right)} = \half$, confirming our earlier intuition.

From a graph-blind RIC's perspective, these numerical investigations not only validate the optimality of fully-biased BIPs and their asymptotic gains over unbiased BIPs, but also reveal that: (i) even partially-biased BIPs lose substantially against the fully- and similarly-biased ones, and (ii) the asymptotic performance is approached quite rapidly for increasing $n$.

 
 
 




\section{Conclusions} \label{sec:conclusion}

We investigated an information fusion problem in a network, primarily motivated by the unique qualities and constraints of modern-day social networks. Herein, information-providers supply only binary information, are unrestrained, prone to bias, and abundantly available. In such an information network comprised of biased-information-providers (BIPs) and rational-information-consumers (RICs), we studied the impact of the information-providers' biases (tendencies of favoring one outcome over the other) on a graph-blind RIC that is trying to deduce the true information.

We modeled the input-output behavior of each BIP as a binary channel, and reasonably assumed that an RIC (especially, a human one) might not know/account for the dependencies among its BIPs. Then, we proved that from this ``graph-blind" RIC's perspective, the optimal decision will always be to choose BIPs that are fully-biased against the a-priori likely event. \emph{Further, in the absence of BIPs that are fully-biased against the a-priori likely event, the set of BIPs that the RIC will choose are still guaranteed to be maximally-biased.} We also proved that by choosing $n$ identical, fully-biased BIPs instead of $n$ identical, unbiased BIPs, the RIC anticipates gains that converge asymptotically to an exponential growth rate with a positive exponent. We explicitly characterized this exponent using the number of BIPs $n,$ their common error rate $r>0$, and the prior distribution of the input at the source $\rho_0.$
%

Our work establishes that, \emph{if a graph-blind RIC is accounting for bias}, it will choose its information-providers according to a counter-intuitive, but highly tractable rule: ``When choosing $n$ BIPs from infinitely many, (presumably) independent BIPs reporting a $0/1$ event, choose the ones that are fully-biased against the a-priori likely event." This rule is opposed to the more intuitive option of choosing diversely-biased or unbiased information-providers.

Any information-consumer can emulate an RIC by following a counter-intuitive, but elegantly uniform rule. However, if the BIPs are not acting independently, then the dependencies might affect the accuracy of the information that an RIC deduces to be true. In accordance to the varying levels of network awareness that might be exhibited by rational agents with limited resources (e.g. humans, low-complexity devices), in future work, we hope to build on this model of the graph-blind RIC to arrive at models of ``more-aware" RICs. Through these models, we aim to enable the design of novel RIC-centric frameworks with mechanisms to reduce misinformation.




\bibliographystyle{IEEEtran}
\bibliography{ConcavityOfErrorWrtBias}



\end{document}